\documentclass[epj]{svjour}
\usepackage{xspace}
\usepackage{subfig}
\usepackage [dvips]{graphics, color}

\newcommand{\sNNtwohundred}{$\sqrt{s_{NN}}$ = 200 GeV\xspace}
\newcommand{\sNNsixtytwo}{$\sqrt{s_{NN}}$ = 62.4 GeV\xspace}
\newcommand{\sNN}{$\sqrt{s_{NN}}$\xspace}
\newcommand{\stdassoc}{$1.5$ GeV/c $<$ $p_T^{associated}$ $<$ $p_T^{trigger}$\xspace}
\newcommand{\stdtrig}{$3.0$ $<$ $p_T^{trigger}$ $<$ $6.0$ GeV/c\xspace}
\newcommand{\pttrig}{$p_T^{trigger}$\xspace}
\newcommand{\ptassoc}{$p_T^{associated}$\xspace}
\newcommand{\npart}{$N_{part}$\xspace}
\newcommand{\pT}{$p_T$\xspace}
\newcommand{\highpT}{high-$p_T$\xspace}
\newcommand{\jet}{\textit{Jet}\xspace}
\newcommand{\njet}{$Y_{Jet}$\xspace}
\newcommand{\ridge}{\textit{Ridge}\xspace}
\newcommand{\nridge}{$Y_{Ridge}$\xspace}

\newcommand{\pp}{$p+p$\xspace}
\newcommand{\Cu}{$Cu+Cu$\xspace}
\newcommand{\Au}{$Au+Au$\xspace}
\newcommand{\A}{$A+A$\xspace}
\newcommand{\dAu}{$d+Au$\xspace}

\newcommand{\zT}{$z_T$\xspace}
\newcommand{\MeV}{MeV/c\xspace}
\newcommand{\GeV}{GeV/c\xspace}
\newcommand{\dphi}{$\Delta\phi$\xspace}
\newcommand{\deta}{$\Delta\eta$\xspace}
\newcommand{\vtwo}{$v_2$\xspace}
\newcommand{\pthat}{$\hat{p_T}$\xspace}
\newcommand{\nearside}{near-side\xspace}

\newcommand{\fref}[1]{Fig.~\ref{#1}}
\newcommand{\tref}[1]{Tab.~\ref{#1}}
\newcommand{\Fref}[1]{Fig.~\ref{#1}}

\newcommand{\etal}{et al.}

\begin{document}
\title{Energy and system dependence of \highpT triggered two-particle near-side correlations}
\author{Christine Nattrass\inst{1} for the STAR collaboration
}                     
%
%
\institute{WNSL, 272 Whitney Ave., Yale University, New Haven, CT 06520, USA}
\authorrunning{Nattrass (STAR)}
\titlerunning{Energy and system dependence of two-particle correlations}
\date{Received: date / Revised version: date}
%
\abstract{
Previous studies have indicated that the near-side peak of \highpT triggered correlations can be decomposed into two parts, the \jet and the \ridge.  We present data on the yield per trigger of the \jet and the \ridge from \dAu, \Cu and \Au collisions at \sNNsixtytwo and 200 GeV and compare data on the \jet to PYTHIA 8.1 simulations for \pp.  PYTHIA describes the \jet component up to a scaling factor, meaning that PYTHIA can provide a better understanding of the \ridge by giving insight into the effects of the kinematic cuts.
We present collision energy and system dependence of the \ridge yield, which should help distinguish models for the production mechanism of the \ridge.
} 
\maketitle
\section{Introduction}\label{Introduction}
Previous studies in \Au collisions at \sNNtwohundred demonstrated that the near-side peak in high-\pT triggered correlations can be decomposed into two structures.  The \jet is narrow in both azimuth (\dphi) and pseudorapidity (\deta), similar to what is observed in \dAu, while the \ridge is narrow in azimuth but broad in pseudorapidity.  The \jet component is similar to that expected from vacuum fragmentation, whereas the \ridge has properties similar to the bulk \cite{Jana,Joern}.  Comparing data from \Au and \Cu collisions at \sNNsixtytwo and \sNNtwohundred tests whether these conclusions hold for other collision systems and energies.

Several mechanisms have been proposed for the production of the \ridge \cite{LongFlow,MomentumKick,Reco,Sergei,PlasmaInstability}.  These models have yielded few calculations which can be directly compared to data, in part because of the large number of factors which must be considered when theoretically calculating the experimentally measured quantitites.  The results presented here should provide a good test of models for the production of the \jet and \ridge because trends expected with changing collision energy and in nuclei collided in a given model should be easier to calculate theoretically.

\section{Method}\label{Method}
Data from the STAR detector from year 3 \dAu collisions as \sNNtwohundred, year 4 \Au collisions at \sNNsixtytwo and \sNNtwohundred, and year 5 \Cu collisions at \sNNsixtytwo and \sNNtwohundred were used for the comparison of collision systems and energies.  Details of the STAR detector can be found in \cite{STARNIM}.  The primary detector used for these analyses was the STAR Time Projection Chamber (TPC.)

A high transverse momentum (\pT) particle is selected and the distribution of other particles in the event relative to that trigger particle in azimuth (\dphi) and pseudorapidity (\deta) $\frac{d^2N}{d\Delta\phi d\Delta\eta}$ was determined.  The \pT of the trigger and associated particles was restricted in order to reduce the soft background; unless otherwise mentioned \stdassoc and \stdtrig.  $\frac{d^2N}{d\Delta\phi d\Delta\eta}$ is normalized by the number of trigger particles.
This was corrected for the single particle efficiency and for detector acceptance, which is dependent on the collision system and energy, \pT, \deta, \dphi, and collision multiplicity.  Except for studies of \npart dependence, the \Cu data at both energies are for 0-60\% centrality, \Au data at \sNNsixtytwo are for 0-80\% centrality, and \Au data at \sNNtwohundred are for 0-10\% centrality.  \dAu data are minimum bias.

The yield measured is the number of particles associated with the trigger particle within limits on \ptassoc and \pttrig.  The \ridge was previously observed to be roughly independent of \deta within the acceptance of the STAR TPC \cite{Joern}.  To extract the yield it is assumed that the \ridge is independent of \deta.  Previous studies have demonstrated that the \jet component extends to $|$\deta$|$ = 0.75 in the \pT range studied here and that limited detector acceptance limits studies to $|$\deta$|$ $<$1.75 \cite{Jana,Joern,MeSQM}.  To determine the \jet yield \njet, the projection of the distribution of particles $\frac{d^2N}{d\Delta\phi d\Delta\eta}$ is taken in two different ranges in pseudorapidity:

$\frac{dY_{Ridge}}{d\Delta\phi}$ = $1/N_{trigger} \int\limits_{-1.75}^{-0.75} \frac{d^2N}{d\Delta\phi d\Delta\eta} d\Delta\eta$ 

\hspace{1cm}+ $1/N_{trigger} \int\limits_{0.75}^{1.75} \frac{d^2N}{d\Delta\phi d\Delta\eta} d\Delta\eta$

$\frac{dY_{Jet+Ridge}}{d\Delta\phi}$ = $1/N_{trigger} \int\limits_{-0.75}^{0.75} \frac{d^2N}{d\Delta\phi d\Delta\eta} d\Delta\eta$

where the former contains only the \ridge and the latter contains both the \jet and the \ridge.  The jet-like yield on the \nearside is the integral over $-1<$ \dphi $<1$:

$Y_{Jet}$ = $\int\limits_{-1}^{1} ( \frac{dY_{Jet+Ridge}}{d\Delta\phi} - \frac{0.75}{1}  \frac{dY_{Ridge}}{d\Delta\phi} ) d\Delta\phi$.

The factor in front of the second term is the ratio of the \deta width in the region containing the \jet and the \ridge to the width of the region containing only the \ridge.  With this method for subtracting the \ridge contribution to \njet, the systematic errors due to \vtwo cancel out assuming that \vtwo is roughly independent of \deta, a reasonable assumption in the mid-rapidity range $|\eta| < 1$ based on the available data \cite{phobosFlow1,phobosFlow2}.  It is also assumed that the \ridge is independent of \deta.

To determine \nridge the integration is done over the entire \deta region to minimize the effects of statistical fluctuations in the determination of the background:

$Y_{Ridge}$ = $1/N_{trigger} \int\limits_{-1.75}^{1.75} \int\limits_{-1}^{1} \frac{d^2N}{d\Delta\phi d\Delta\eta} d\Delta\phi d\Delta\eta$ - $Y_{Jet}$.

The integration over \dphi is done by fitting a Gaussian to the \nearside.  This partially compensates for a detector effect which causes lost tracks at \dphi $\approx$ 0 and \deta $\approx$ 0; this effect is less than 10\% in the \pT range studied here \cite{Marek}.

The raw signal has a background due to particles correlated indirectly with each other in azimuth due to their correlation with the reaction plane.  This random background is given by

$\frac{dY_{bkgd}}{d\phi} = B (1 + 2 \langle v_2^{trigger}\rangle \langle v_2^{associated}\rangle \cos(2\Delta\phi) )$ 

where \vtwo is the second order harmonic in a Fourier expansion of the momentum anisotropy relative to the reaction plane, and must be subtracted in order to study the component associated with the jet.  Systematic errors come from the errors on B, $\langle v_2^{trigger}\rangle$ and $\langle v_2^{associated}\rangle$. It is assumed that \vtwo is the same for events with a trigger particle as for minimum bias events and that \vtwo is roughly independent of \deta.  For each data set $v_2(p_T)$ was fit in centrality bins to determine $\langle v_2^{trigger}\rangle$ and $\langle v_2^{associated}\rangle$.  Details of the \vtwo subtraction for \Au collisions at \sNNtwohundred are given in \cite{Jana} and for \Cu collisions at \sNNtwohundred in \cite{MeSQM}.  For \Cu collisions at \sNNsixtytwo, the \vtwo using the reaction plane as determined from tracks in the Forward Time Projection Chamber was used as the nominal value and the lower bound was determined from a multiplicity-dependent approximation as described for \sNNtwohundred in \cite{MeSQM}.  For \Au collisions at \sNNsixtytwo, \vtwo and its systematic errors were taken from \cite{AuAuSixtyTwoFlow}.  B is fixed using the ZYAM method \cite{starZYAM}.


PYTHIA 8.1 was used to simulate \pp collisions for comparisons to \njet.  A trigger particle was selected and the distribution of particles in azimuth was calculated, as in the experimental measurements.  The yield was determined as the number of charged hadrons in the range $-1<$ \dphi $<1$.  For comparisons to data identical limits on \ptassoc and \pttrig were applied.  The minimum \pthat is the parameter in PYTHIA for the transverse momentum in the hard subprocess \cite{PYTHIAManual}.  A minimum value of \pthat = 0.1 \GeV was used and $10^8$ events were simulated to ensure that the minimum \pthat did not affect the yield and that the statistical error was negligible.   It was not necessary to study the distribution of particles in pseudorapidity since there is no \ridge in PYTHIA.

\section{Results}\label{Results}
\subsection{The \jet}\label{TheJet}
\begin{figure}
\resizebox{8.8cm}{!}{%
  \includegraphics{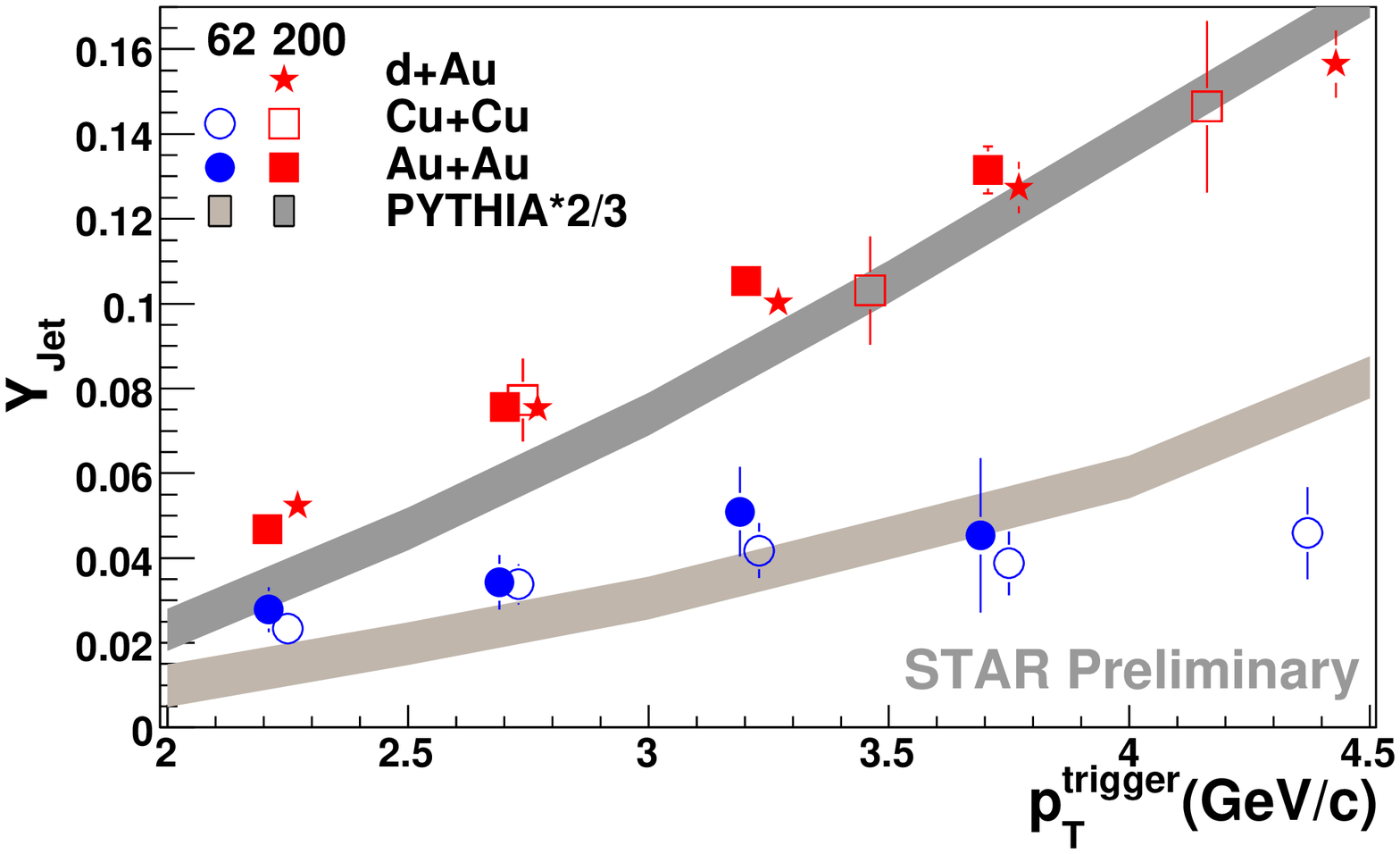}
}
\caption{\pttrig dependence of the \njet for \Cu and \Au at \sNNsixtytwo and \dAu, \Cu, and \Au at \sNNtwohundred compared to the yield from PYTHIA scaled by 2/3.  Color online.}
\label{TrigPt}       
\end{figure}

\Fref{TrigPt} compares the dependence of \njet on \pttrig for all systems and energies to the yield from PYTHIA 8.1 scaled by 2/3.  An overall scaling factor of 2/3 was applied to the PYTHIA yields to match the data.  
The need for the scaling factor implies that PYTHIA assumes that too many particles are produced in hard processes, however, kinematic effects should still be reflected accurately in PYTHIA.
The scaled PYTHIA yield describes the shape of the \pttrig dependence well, with a few deviations at lower \pttrig.  PYTHIA describes the energy dependence of \njet well, indicating that the energy dependence can be explained as a pQCD effect.  If \njet is dominated by pQCD effects, deviations from PYTHIA at lower \pT would be expected.  No system dependence is observed in the data, as would be expected for an effect dominated by pQCD.

\begin{figure}
\resizebox{8.8cm}{!}{%
  \includegraphics{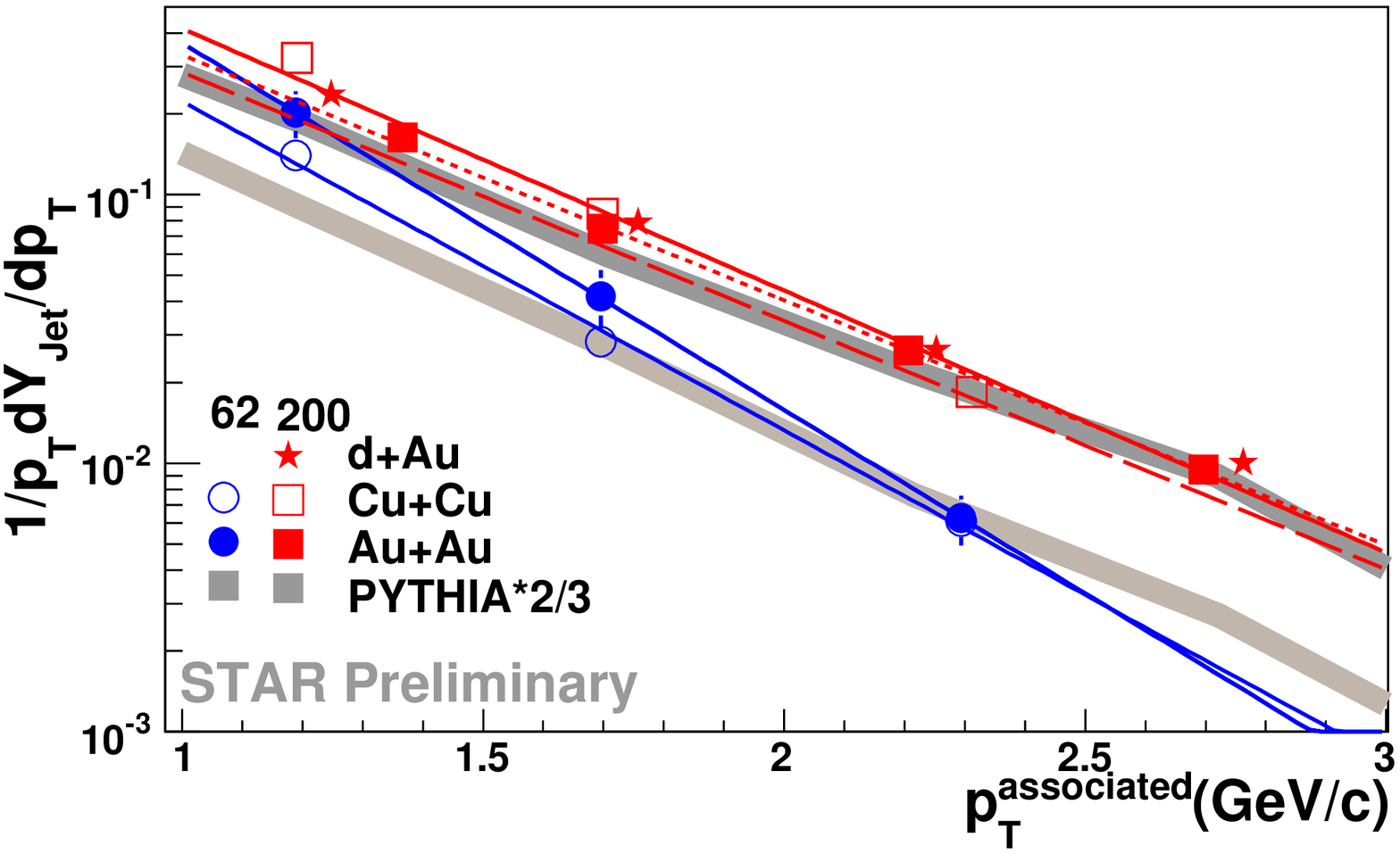}
}
\caption{\ptassoc dependence of \njet for \Cu and \Au at \sNNsixtytwo and \dAu, \Cu, and \Au at \sNNtwohundred compared to the yield from PYTHIA scaled by 2/3.  The inverse slope parameters from fits of an exponential to the data and to PYTHIA are given in \tref{Table}.  Color online.}
\label{AssocPt}       
\end{figure}

The dependence of \njet on \ptassoc is shown in \fref{AssocPt}.  As in \fref{TrigPt}, the scaled PYTHIA yield describes the shape of the data well and there is no system dependence.  The inverse slope parameters from exponential fits to the data and to PYTHIA shown in \tref{Table} likewise support independence on collision system.   Slight deviations from the scaled PYTHIA yield at lower \ptassoc in \fref{AssocPt} for collisions at \sNNsixtytwo are reflected in the inverse slope parameter, which is higher than that of the data.
\begin{table}
\caption{Inverse slope parameter k (\MeV) of \ptassoc for fits of data in \Fref{AssocPt}.  The inverse slope parameter from a fit to $\pi^-$ in \Au from \cite{Pion} above 1.0 \GeV is $k$ = 280.9 $\pm$ 0.4 \MeV for \sNNsixtytwo and is $k$ = 330.9 $\pm$ 0.3 \MeV for \sNNtwohundred.  Statistical errors only.}\label{Table}

\begin{tabular}[b]{c|c|c}
\hline
 & \sNNsixtytwo & \sNNtwohundred\\
   & h-h & h-h\\ \hline
\Au \jet & 317 $\pm$ 26 & 478 $\pm$ 8 \\
\Cu \jet & 355 $\pm$ 21 & 445 $\pm$ 20 \\ 
\dAu \jet &  & 469 $\pm$ 8 \\ \hline
PYTHIA & 424 $\pm$ 5 & 473 $\pm$ 3\\ \hline
\end{tabular}
\end{table}
\begin{figure}
\resizebox{8.8cm}{!}{%
  \includegraphics{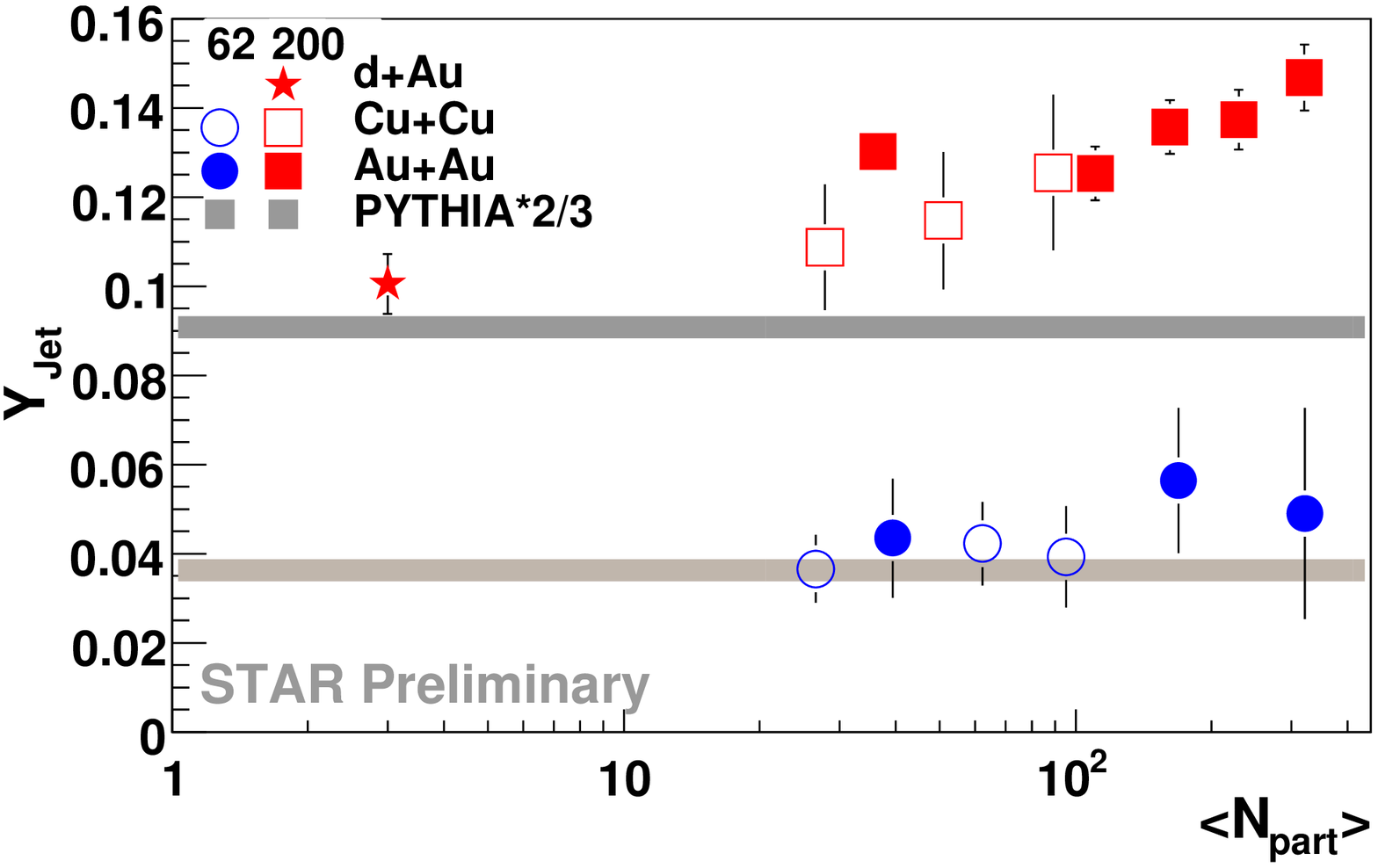}
}
\caption{\npart dependence of the \njet for \Cu and \Au at \sNNsixtytwo and \dAu, \Cu, and \Au at \sNNtwohundred compared to the yield from PYTHIA.  Color online.}
\label{JetNpart}       
\end{figure}

The \npart dependence of \njet is shown in \fref{JetNpart} and compared to the scaled PYTHIA yield.  In contrast to measurements at higher \pT,g which show no \npart depedence, a small increase with \npart
is observed.  This may be caused by either slight modifications of the \jet which increase with system size or  some of the \ridge being misidentified as part of the \jet.  If the \ridge were not completely independent of \deta, some of the particles in the \ridge could be associated with the \jet.  Since the \ridge has roughly four times as many particles than the \jet in central \Au, this would give a smaller relative error to the \ridge than the \jet.  However, the \jet has also been observed to be considerably broader in \deta in \A collisions than in \pp and \dAu collisions \cite{Joern,Marek}, which would imply modifications of the \jet in \A collisions.  At this point models for \jet and \ridge production cannot distinguish these mechanisms.


\begin{figure}
\resizebox{8.8cm}{!}{%
  \includegraphics{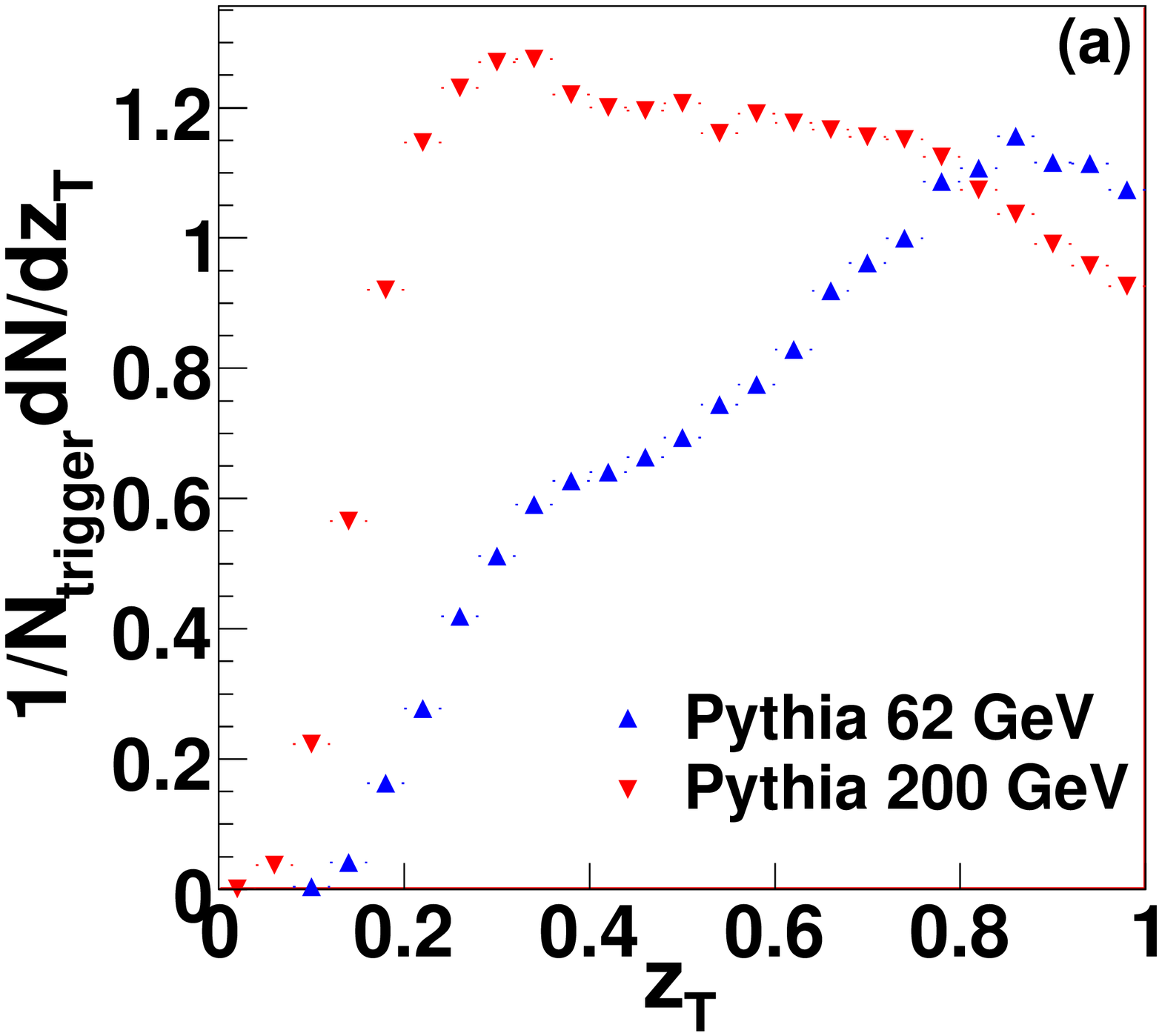}
  \includegraphics{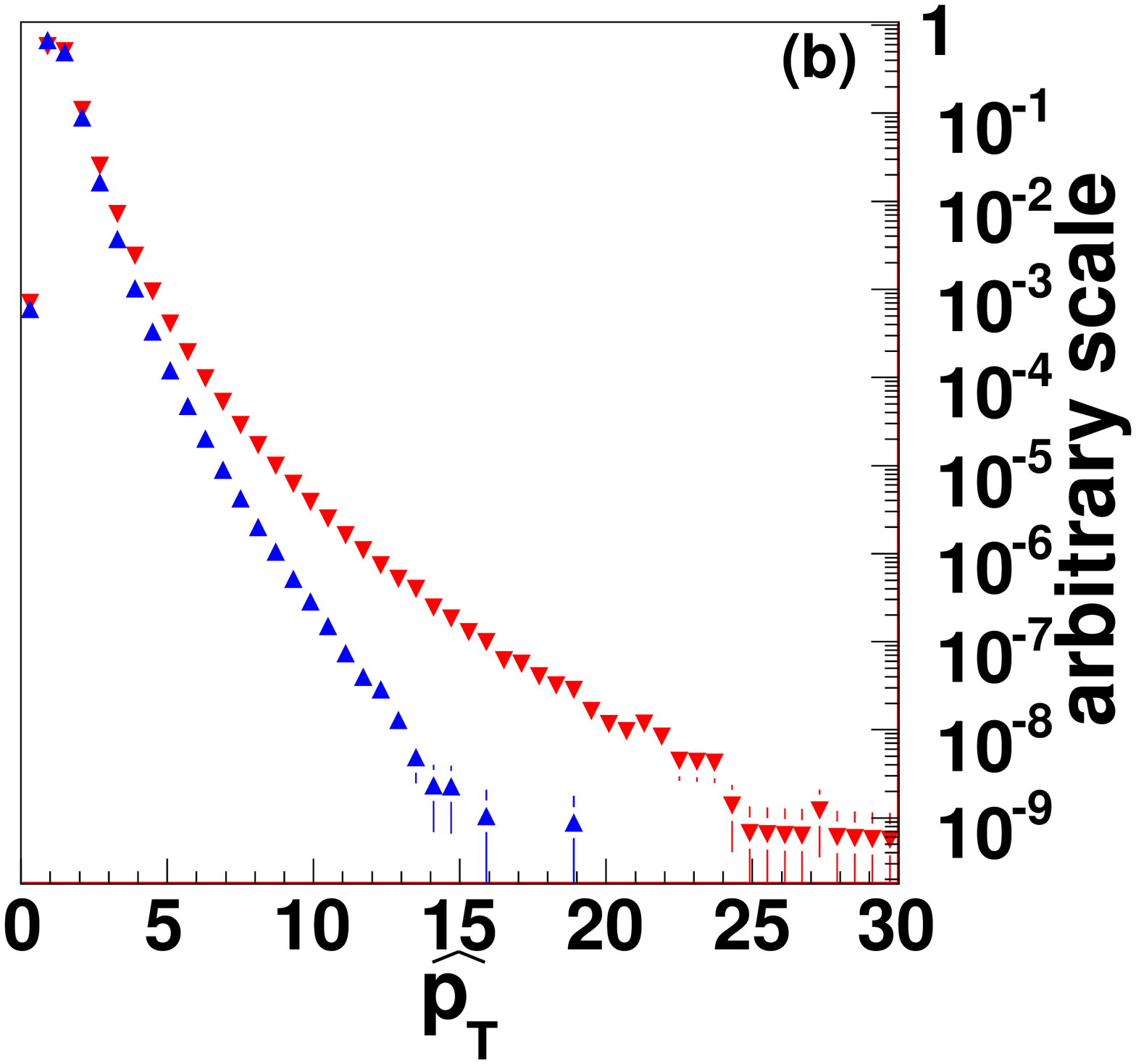}
}
\caption{(a) distribution of trigger particles \zT and (b) \pthat distribution in PYTHIA at \sNNsixtytwo from PYTHIA 8.1 for \stdassoc and \stdtrig at \sNNsixtytwo and \sNNtwohundred.  Color online.}
\label{PYTHIA}       
\end{figure}
That PYTHIA describes the \pttrig and \ptassoc dependence of \njet fairly well implies that PYTHIA can be used to approximate the momentum fraction carried by the leading hadron, \zT.  \Fref{PYTHIA} shows the \pthat distribution and the distribution of trigger particles in \zT = \pttrig/\pthat predicted by PYTHIA for \sNNsixtytwo and \sNNtwohundred.  \Fref{PYTHIA}(a) shows that for the same \pttrig and \ptassoc, the mean \zT is higher in \sNNsixtytwo and therefore the mean jet energy is lower.  \Fref{PYTHIA}(b) shows that this is caused by the steeper spectrum at \sNNsixtytwo.  The lower \njet in collisions at \sNNsixtytwo results from the higher mean \zT and is a kinematic effect.

\subsection{The \ridge}\label{TheRidge}

\begin{figure}
\resizebox{8.8cm}{!}{%
  \includegraphics{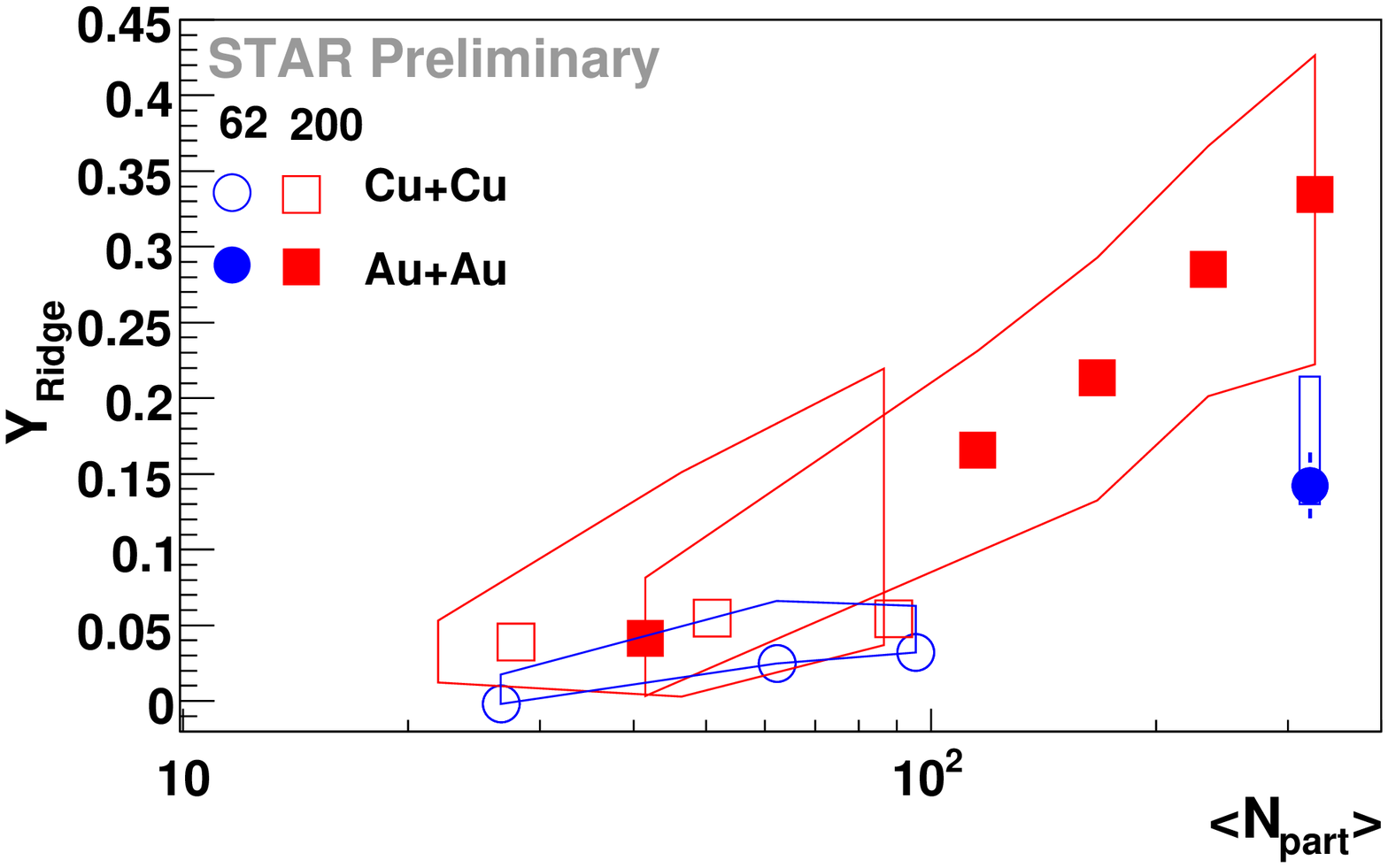}
}
\caption{\nridge dependence on \npart for \sNNsixtytwo and\sNNtwohundred.  Color online.}
\label{RidgeNpart}       
\end{figure}

\begin{figure}
\resizebox{8.8cm}{!}{%
  \includegraphics{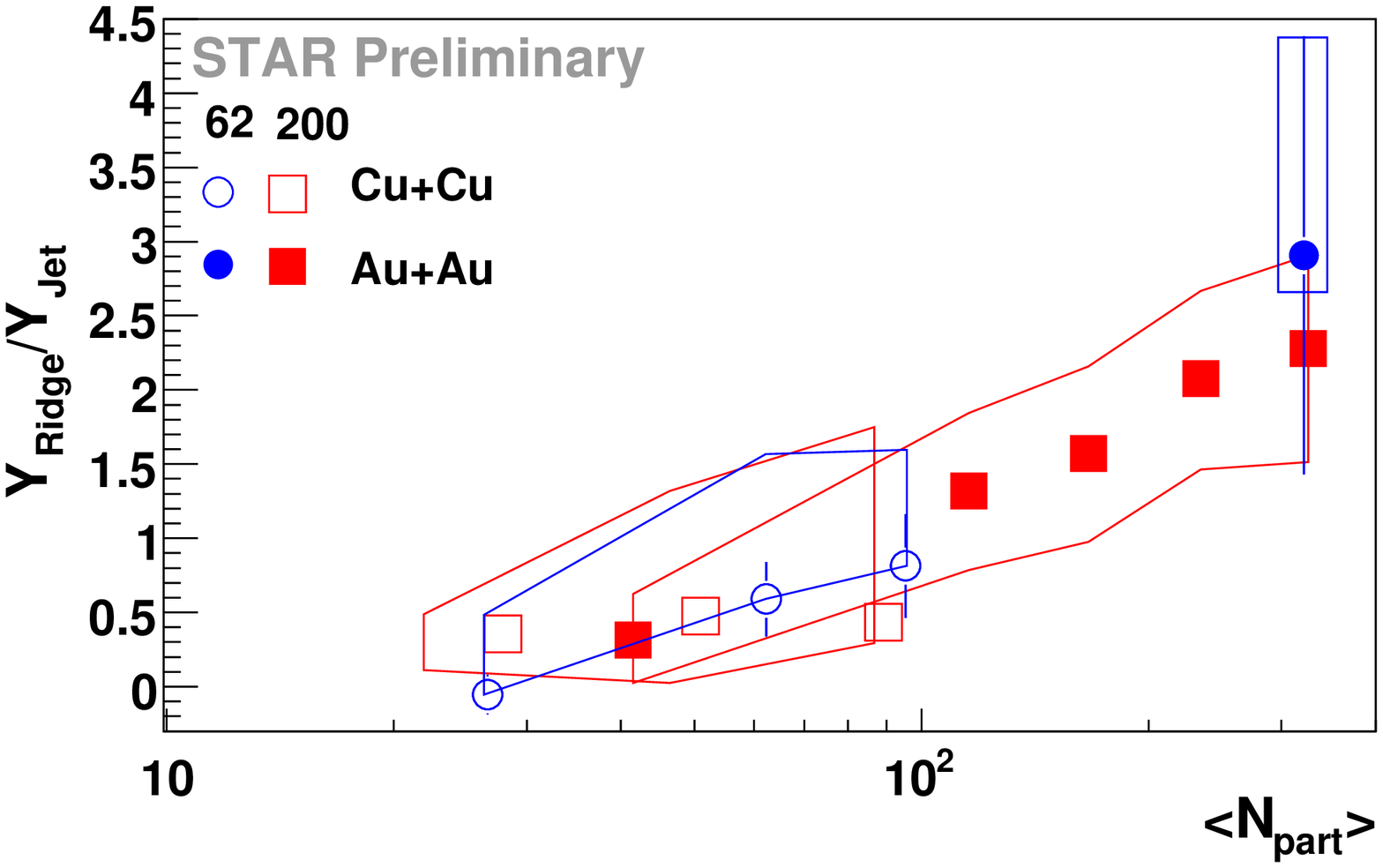}
}
\caption{\nridge/\njet dependence on \npart for \sNNsixtytwo and\sNNtwohundred.  Color online.}
\label{RidgeRatio}       
\end{figure}

The dependence of \nridge on \npart is given in \fref{RidgeNpart}.  In collisions at both \sNNsixtytwo and \sNNtwohundred \nridge increases with \npart.  As seen in \fref{JetNpart}, the yield at \sNNsixtytwo is considerably smaller than at \sNNtwohundred.  \Fref{RidgeRatio} shows the ratio \nridge/\njet and shows that this ratio does not depend on \sNN.  PYTHIA simulations demonstrated that the data at \sNNsixtytwo likely correspond to a lower jet energy, so this implies that \nridge decreases with energy just like \njet.

Few models have attempted to make quantitative predictions for \nridge.  An exception is the momentum kick model, which is consistent with data on the energy dependence of \nridge \cite{WongEnergy}.  The collision energy dependence of \nridge is potentially a sensitive test of models because the dominant factor in collision energy dependence should be different for various classes of models.  Models which involve parton energy loss due to interaction with the medium such as the momentum kick model should have a smaller \ridge at lower energy, as observed in the data, because the initial parton energy was lower.  The radial flow+trigger bias model should predict a dependence of \nridge on the amount of radial flow in the system.  An analysis similar to \cite{SergeiLatest} could yield predictions for the collision system and energy dependence.  Plasma instability models should depend on whether plasma instabilities are more or less likely in small systems and at lower energies.  When more detailed calculations are available, it is likely that the data could exclude some production mechanisms.

\section{Conclusions}
The data from \dAu, \Cu, and \Au and \sNNsixtytwo and \sNNtwohundred demonstrate that the \jet shows no system dependence.  In addition, the collision energy dependence of \njet is described well by PYTHIA even at fairly low \pT and the \pttrig and \ptassoc dependencies agree with PYTHIA up to a scaling factor, with a few deviations at lower \pT.  This implies that the dominant production mechanism of the \jet is fragmentation.  Deviations from PYTHIA may imply modifications of the \jet in \A collisions.  It also implies that PYTHIA or other models can be used to determine the effect of the kinematic cuts on \pttrig on the \zT and jet energy distribution, which could be very useful for the theoretical interpretation of the \ridge.  

\nridge is smaller at lower collision energies and increases with system size indepent of collision system.  There is no dependence on the collision system.  Data on the collision energy and system dependence could provide a robust test of models, and comparisons of \njet to PYTHIA imply that the effects of the kinematic cuts on the distribution of jet energies can be inferred from PYTHIA.
%

\end{document}